\documentclass[12pt]{elsarticle}
\usepackage{fullpage}
\usepackage{amsmath,amsfonts}
\usepackage[ruled,vlined,linesnumbered]{algorithm2e}
\usepackage{xcolor}
\usepackage{hyperref}
\usepackage{url}
\usepackage[section]{placeins}

\usepackage{pgfplots}
\pgfplotsset{compat=newest}

\usetikzlibrary{plotmarks}
\usetikzlibrary{arrows.meta}
\usepgfplotslibrary{patchplots}
\usepackage{grffile}

\usepackage{listings}
\definecolor{lightblue}{rgb}{0.68, 0.85, 0.9}
\definecolor{iceberg}{rgb}{0.44, 0.65, 0.82}

\usepackage{listings}

\usepackage{hhline}

\newcommand{\RR}{\mathbb{R}}
\newcommand{\OO}{\mathcal{O}}
\newcommand{\diag}{\text{diag}}
\newcommand{\ee}{\mathrm{e}}

\begin{document}

\title{Efficient 6D Vlasov simulation using the dynamical low-rank framework \texttt{Ensign}}
\author[trento]{Fabio Cassini\corref{cor1}} \ead{fabio.cassini@unitn.it}
\author[uibk]{Lukas Einkemmer} \ead{lukas.einkemmer@uibk.ac.at}
\address[trento]{Department of Mathematics, University of Trento, Italy}
\address[uibk]{Department of Mathematics, University of Innsbruck, Austria}
\cortext[cor1]{Corresponding author}

\begin{abstract} Running kinetic simulations using grid-based methods is
extremely expensive due to the up to six-dimensional phase space. Recently, it
has been shown that dynamical low-rank algorithms can drastically reduce the
required computational effort, while still accurately resolving important
physical features such as filamentation and Landau damping.
In this paper, {\color{black}we propose a new second order projector-splitting 
dynamical low-rank algorithm for the full six-dimensional 
Vlasov--Poisson equations.
An exponential integrator based Fourier spectral method is
employed to obtain a numerical scheme that is CFL condition free but still fully
explicit.
The resulting method is implemented with the aid of \texttt{Ensign}, a software framework
which facilitates the efficient implementation of dynamical low-rank algorithms
on modern multi-core CPU as well as GPU based systems.
Its usage and features are briefly described
in the paper as well.}
The presented numerical results demonstrate that 6D simulations can
be run on a single workstation and highlight the significant speedup
that can be obtained using GPUs.
\end{abstract}
\maketitle

\section{Introduction}

Efficiently solving kinetic equations is important in applications
ranging from plasma physics to radiative transport. The main challenge
in this context is the up to six-dimensional phase space and the associated
unfavorable scaling of computational cost and memory requirements. Assuming $n$
discretization points for each direction of a 6D phase space, the storage cost
of a  direct discretization scales as $\OO(n^{6})$. This is usually referred
to as the curse of dimensionality.
To mitigate this issue, many techniques have been proposed in the literature;
we mention, for example, particle methods~\cite{camporeale2016,verboncoeur2005}
and sparse grid approximations~\cite{guo2016,kormann2016}. However, it is well
known that the former misses or underresolves some important physical phenomena
(such as Landau damping or regions with low density), while the latter has
issues with the Gaussian equilibrium distribution and the low regularity
inherent in collisionless kinetic problems. Because of this, direct simulations
are routinely conducted on large supercomputers~\cite{bigot2013scaling,einkemmer2016high}.
However, current
computational constraints mostly limits this to four- and some five-dimensional
problems.

More recently, a dynamical low-rank algorithm for solving the
Vlasov--Poisson {\color{black}equations} has been proposed in~\cite{einkemmer2018low}.
{\color{black} In the context of a 6D phase space,} dynamical low-rank integrators
approximate the {\color{black}solution} by a set of only three-dimensional
advection problems. The resulting algorithm has the primary advantage of having
storage and computational costs that scale as {\color{black}$\OO(rn^{3})$ and 
$\OO(r^2n^3)$, respectively,} where $r$ is the
(usually small) rank of the approximation. This can result in a drastic
reduction of both memory consumption as well as computational effort.
The main enabling technology was the introduction of the
projector-splitting integrator~\cite{lubich2014projector}, which allows us to
obtain a robust integrator without the need for regularization within the
framework of dynamical low-rank approximations~\cite{Koch2007,Lubich2008,meyer09mqd,meyer90tmc}.

For kinetic equations the dynamical low-rank approach offers a range of
advantages, {\color{black}not only strictly related to storage and computational costs.
For example,} filamented structures in velocity space can be resolved
accurately~\cite{einkemmer2020vlasovmaxwell}.
Moreover, it is known that in the linear regime~\cite{einkemmer2018low} and
for certain fluid~\cite{einkemmer2018compr,einkemmer2021bgk} and {\color{black}diffusive
limits}~\cite{ding2021,einkemmer2021b} the solution has a low-rank structure,
{\color{black} hence it is natural to use a low-rank approximation}.
Recently, a dynamical algorithm that is conservative from first principle has
been constructed~\cite{einkemmer2021mass}. Because of these advances,
dynamical low-rank approximations have received significant interest lately,
and methods for problems from plasma physics~\cite{einkemmer2018low,einkemmer2020vlasovmaxwell,guo2021low,kormann2015semi},
radiation transport~\cite{einkemmer2021b,kusch2021stability,peng2021high,peng2020low},
and uncertainty quantification for hyperbolic problems~\cite{kusch2021dynamical}
have been proposed. These schemes have the potential to enable the 6D simulation
of such systems on small clusters or even desktop computers.

While mature software packages exist for solving kinetic problems using both
particle methods (e.g.~\cite{burau2010picongpu,tskhakaya2007optimization})
and methods that directly discretize the phase space on a grid
(e.g.~\cite{einkemmer2019,grandgirard2006gysela,selalib,von2014vlasiator}),
no such software frameworks exist for dynamical low-rank algorithms.
However, the need in the latter case is arguably even more critical, as the
resulting evolution equations are usually somewhat more complex than the
original model.

The purpose of this paper is to present {\color{black} six-dimensional simulations
of the Vlasov--Poisson equations using} the framework \texttt{Ensign}%
\footnote{{\color{black}Publicly
available at \url{https://github.com/leinkemmer/Ensign} under the MIT license}},
which facilitates the easy and efficient implementation of dynamical low-rank
algorithms for kinetic equations (both on multi-core CPU and GPU based systems).
{\color{black}In particular, we will employ a newly designed
second order} projector-splitting dynamical low-rank algorithm,
which is based on a CFL-free (but still fully explicit)
{\color{black} exponential integrator } that uses Fourier spectral methods {\color{black}to compute the action of certain matrix functions}.
We emphasize, however, that {\color{black} in principle} our software framework
is able to support other dynamical low-rank techniques (such as the unconventional
integrator~\cite{ceruti2021unconventional}), and is completely flexible with
{\color{black}regard} to the specific space and time discretizations employed.
{\color{black} In fact, almost any third party library suitable for a given
problem could be used alongside \texttt{Ensign} in an implementation}.

The remainder of the paper is structured as follows:
{\color{black} in Section~\ref{sec:lrcont} we describe the general structure of
a projector-splitting dynamical low-rank integrator for the Vlasov--Poisson
equations. In Section~\ref{sec:matform} we then semi-discretize (discrete in space and continuous in time)
the equations of motion resulting from the projector-splitting integrator
and write them in a matrix formulation. The proposed integrator is described in
detail in Section \ref{sec:tintKSL}. In Section~\ref{sec:usage} we provide a
big picture overview of the \texttt{Ensign} software framework and
how it can be used to implement a dynamical low-rank algorithm. We present some numerical
results in Section~\ref{sec:numerics} and discuss the performance of the
dynamical low-rank algorithms on CPUs and GPUs in Section~\ref{sec:performance}.
Finally, we conclude in Section~\ref{sec:conc}.}

\section{Low-rank approximation for the Vlasov--Poisson {\color{black}equations}} \label{sec:lrcont}

In this paper we consider the Vlasov--Poisson {\color{black}equations} in
dimensionless form given by
\begin{equation} \label{eq:vp}
  \left\{\begin{aligned}
    &\partial_t f(t,x,v) + v \cdot \nabla_x f(t,x,v) - E(f)(t,x) \cdot \nabla_v f(t,x,v) = 0, \\
    &E(f)(t,x) = -\nabla_x \phi(t,x){\color{black},} \\
    & -\Delta\phi(t,x) = \rho(f)(t,x) + 1, \quad \rho(f)(t,x) = -\int_{\Omega_v}f(t,x,v) \,dv,
  \end{aligned}
  \right.
\end{equation}
where $f(t,x,v)$ represents the particle-density function of the species under
consideration, $t\in \RR_0^+$ is the time variable, $x\in\Omega_x \subset \RR^d$
refers to the space variable, $v\in\Omega_v \subset \RR^d$ is the velocity
variable and $d=1,2,3$. Depending on the physical phenomenon under study,
system~\eqref{eq:vp} is completed with appropriate boundary and initial conditions.

We now describe how to obtain the dynamical low-rank
approximation of the Vlasov--Poisson system~\eqref{eq:vp}. 
{\color{black}A reader not familiar with these concepts can find more details,
for instance, in~\cite{einkemmer2018low}}.
The computational domain is denoted by $\Omega = \Omega_x \times \Omega_v$.
Then, instead of {\color{black}directly solving system}~\eqref{eq:vp}, we look for an
approximation of the particle-density function $f(t,x,v)$ that, for fixed $t$,
lies in the rank-$r$ manifold
  \begin{equation*}
  \begin{aligned}
    \mathcal{M} =
    \Big\{g(x,v) \in L^2(\Omega) : g(x,v) = \sum_{i,j}X_i(x)S_{ij}V_j(v)
    \text{ with invertible } S=(S_{ij})\in\RR^{r\times r}, \\
    X_i \in L^2(\Omega_x), V_j \in L^2(\Omega_v) \text{ with }
    \langle X_i,X_k\rangle_x =\delta_{ik}, \langle V_j,V_\ell\rangle_v =\delta_{j\ell} \Big\}
  \end{aligned}
\end{equation*}
with corresponding tangent space
\begin{equation*}
  \begin{aligned}
    \mathcal{T}_f\mathcal{M} =
    \Big\{\dot{g}(x,v) \in L^2(\Omega) : \dot{g}(x,v) = \sum_{i,j}\big(X_i(x)\dot{S}_{ij}V_j(v)
    + \dot{X}_i(x)S_{ij}V_j(v) + X_i(x)S_{ij}\dot{V}_j(v)\big), \\
    \text{with } \dot{S} \in \RR^{r\times r},
     \dot{X}_i \in L^2(\Omega_x), \dot{V}_j \in L^2(\Omega_v)
    \text{ and } \langle X_i,\dot{X}_k\rangle_x =0, \langle V_j,\dot{V}_\ell\rangle_v =0 \Big\}.
  \end{aligned}
\end{equation*}
Here $\langle \cdot,\cdot\rangle_x$ and $\langle \cdot,\cdot\rangle_v$ denote
the standard inner products on $L^2(\Omega_x)$ and $L^2(\Omega_v)$, respectively,
and {\color{black} we employ Newton's notation} for the time derivative. Moreover,
all indexes run from $1$ to $r$ and, for simplicity of presentation, we drop
these limits from the notation. 

To obtain the dynamical low-rank approximation of the Vlasov--Poisson
system~\eqref{eq:vp} we need to compute
\begin{equation}\label{eq:tbproj}
  \partial_t f(t,x,v) = - P(f) (v \cdot \nabla_x f(t,x,v) -E(f) \cdot \nabla_v f(t,x,v)),
\end{equation}
where $P(f)$ is the orthogonal projector onto the tangent space
$\mathcal{T}_f\mathcal{M}$. For simplicity of notation, we will {\color{black} keep
using the symbol $f$}
to denote the low-rank approximation to the particle-density.
The orthogonal projection of a generic function $g$ can be written as
\begin{equation*}
  P(f)g = P_{\overline{V}}g - P_{\overline{V}}P_{\overline{X}}g + P_{\overline{X}}g{\color{black},}
\end{equation*}
where $P_{\overline{X}}$ and $P_{\overline{V}}$ are orthogonal projectors onto
the spaces spanned by the functions $X_i$ and $V_j$, respectively. This
formulation suggests a three-term splitting {\color{black}of equation~\eqref{eq:tbproj}}
with {\color{black} subflows (in the sense of differential equations in the context
of splitting schemes) given} by
\begin{align}
  \partial_t f_{\color{black}1}(t,x,v) =
  -P_{\overline{V}}(v \cdot \nabla_x f_{\color{black}1}(t,x,v)
  -E(f_{\color{black}1}) \cdot \nabla_v f_{\color{black}1}(t,x,v)), \label{eq:alKst}\\
  \partial_t f_{\color{black}2}(t,x,v) =
  P_{\overline{V}}P_{\overline{X}}(v \cdot \nabla_x f_{\color{black}2}(t,x,v)
  -E(f_{\color{black}2}) \cdot \nabla_v f_{\color{black}2}(t,x,v)), \label{eq:alSst}\\
  \partial_t f_{\color{black}3}(t,x,v) =
  -P_{\overline{X}}(v \cdot \nabla_x f_{\color{black}3}(t,x,v)
  -E(f_{\color{black}3}) \cdot \nabla_v f_{\color{black}3}(t,x,v)). \label{eq:alLst}
\end{align}

This is the projector-splitting integrator that has been first proposed in~\cite{lubich2014projector}.

By explicitly applying the projector $P_{\overline{V}}$ on equation~\eqref{eq:alKst},
we can see that solving {\color{black} it is equivalent to}
\begin{equation}\label{eq:Kstep}
  \left\{
  \begin{aligned}
    &\partial_tK_j(t,x) = -\sum_{\ell}c_{j\ell}^1\cdot\nabla_xK_\ell(t,x) +
    \sum_{\ell}c_{j\ell}^2\cdot E(K)(t,x)K_\ell(t,x), \\
    &V_j(t,v) =\tilde{V}_j(v),
  \end{aligned}
  \right.
\end{equation}
where the approximate particle-density function is written as
\begin{equation*}
  f_{\color{black}1}(t,x,v) = \sum_{j} K_j(t,x)V_j(t,v),
  \quad K_j(t,x) = \sum_{i} X_i(t,x)S_{ij}(t){\color{black},}
\end{equation*}
and
\begin{equation*}
  c_{j\ell}^1 = \int_{\Omega_v}v\tilde{V}_j(v)\tilde{V}_\ell(v)dv,
  \quad c_{j\ell}^2 = \int_{\Omega_v}\tilde{V}_j(v)\nabla_v\tilde{V}_\ell(v)dv.
\end{equation*}
{\color{black}We refer to~\cite{einkemmer2018low} for a more detailed derivation and a thorough explanation of the underlying mathematical structure.}
Equation~\eqref{eq:Kstep} is usually referred to as the \emph{K step} of the
low-rank projector-splitting algorithm.
Note that the velocity dependent low-rank factors, i.e.~$V_j$, do not change
during this step. {\color{black} This means that} we can use their value at the
beginning of the step, i.e.~$\tilde{V}_j$, in all computations
{\color{black}of the $K$ step}.

By applying both projectors $P_{\overline{V}}$ and $P_{\overline{X}}$, the second
{\color{black} subflow (i.e.~the one related to equation~\eqref{eq:alSst})}, can be obtained by
 \begin{equation}\label{eq:Sstep}
  \left\{
  \begin{aligned}
    &\dot{S}_{ij}(t) = \sum_{k,\ell}(c_{j\ell}^1\cdot d_{ik}^2-c_{j\ell}^2\cdot d_{ik}^1[E(S)])S_{k\ell}(t), \\
    &X_i(t,x)=\breve{X}_i(x), \\
    &V_j(t,v)=\tilde{V}_j(v),
  \end{aligned}
  \right.
 \end{equation}
where
\begin{equation*}
  d_{ik}^1[E(S)] = \int_{\Omega_x}\breve{X}_i(x)E(S)\breve{X}_k(x)dx,
  \quad d_{ik}^2 = \int_{\Omega_x}\breve{X}_i(x)\nabla_x\breve{X}_k(x)dx.
\end{equation*}
We refer to {\color{black} equation}~\eqref{eq:Sstep} as the \emph{S step} of the
low-rank projector-splitting algorithm. Note that neither $X_i$ nor $V_j$
change during this step, {\color{black}i.e.~their values are fixed to the ones
at the beginning of the step ($\breve{X}_i$ and $\tilde{V}_j$, respectively)}.

Finally, we can demonstrate that solving {\color{black} equation}~\eqref{eq:alLst}
is equivalent to
\begin{equation}\label{eq:Lstep}
  \left\{
  \begin{aligned}
    &\partial_tL_i(t,v) = \sum_{k}d_{ik}^1[E(L)]\cdot\nabla_vL_k(t,v) - \sum_{k}(d_{ik}^2\cdot v)L_k(t,v), \\
    &X_i(t,x)=\breve{X}_i(x){\color{black},} \\
  \end{aligned}
  \right.
\end{equation}
with the approximate particle-density function written as 
\begin{equation*}
  f_{\color{black}3}(t,x,v) = \sum_{i} X_i(t,x)L_i(t,v), \quad L_i(t,v) = \sum_{j} S_{ij}(t)V_j(t,v).
\end{equation*}
This step of the low-rank projector-splitting algorithm is referred to as the
\emph{L step}. Note that the $X_i$ do not change during this step,
{\color{black}i.e.~their values are set to $\breve{X}_i$}.

Concerning the equations of the electric field, they can be written in terms of
the low-rank factors $X_i(t,x)$, $S_{ij}(t)$ and $V_j(t,v)$ as well.
Indeed, depending on the need, we can express the charge density $\rho(f)(t,x)$ as
\begin{align*}
  &\rho(K)(t,x) = -\sum_{j} K_j(t,x)\rho(\tilde{V}_j(v)), \qquad \rho(\tilde{V}_j(v))=\int_{\Omega_v}\tilde{V}_j(v) \,dv, \\
  &\rho(S)(t,x) = -\sum_{i,j} \breve{X}_i(x)S_{ij}(t)\rho(\tilde{V}_j(v)), \\
  &\rho(L)(t,x) = -\sum_{i} \breve{X}_i(x)\rho(L_i(t,v)), \qquad \rho(L_i(t,v)) = \int_{\Omega_v}L_i(t,v) \,dv,
\end{align*}
{\color{black} where the relevant quantities are defined above.}

Finally, the approximate solution to the {\color{black}Vlasov--Poisson equations, i.e. system~\eqref{eq:vp}}, is obtained by combining the 
partial solutions of the $K$, $S$ and $L$ step in a splitting fashion.
In the easiest setting, the first order Lie splitting scheme concatenates
the three subflows in sequence, {\color{black}see Section~\ref{sec:foalgo}}.
We will also outline a second order scheme based on Strang splitting
{\color{black} in Section~\ref{sec:soalgo}}.

\section{Matrix formulation of the {\color{black}semi}-discrete algorithm} \label{sec:matform}

So far we have considered the low-rank approximation in a continuous framework.
However, to perform calculations on a computer we have to discretize the
equations for the $K$, $S$, and $L$ steps. In this section, we assume that a
space discretization has been chosen. Our goal is then to collect the degrees
of freedom into matrices and vectors and write the resulting equations
{\color{black} as operations on those objects}.
The implementation using the software framework \texttt{Ensign},
described {\color{black} later in Section~\ref{sec:usage}, is also based on this formulation}.

We consider $n_{x_k}$ discretization points for the space variable $x_k$, 
$k=1,\ldots,d$ and $n_{v_k}$ discretization points for the velocity variable
$v_k$, $k=1,\ldots,d$. The total numbers of degrees of freedom are denoted as
$N_x = n_{x_1}\cdots n_{x_d}$ and $N_v = n_{v_1}\cdots n_{v_d}$ for space and
velocity, respectively.
Then, for a fixed time $t$, we define
$X= [X_1, \dots, X_r]\in \RR^{N_x\times r}$ to be the matrix having
{\color{black} as} columns the evaluation of the low-rank factors $X_i$
{\color{black} at} the chosen spatial grid. Clearly, the resulting matrix entries
depend on the discretization performed and on the ordering of the grid points.
Similarly, we consider $V= [V_1, \dots, V_r]\in \RR^{N_v\times r }$ to be the
matrix having {\color{black} as} columns the evaluation of the low-rank factors
$V_j$ {\color{black} at} the velocity grid. The coupling coefficients $S_{ij}$
are collected in the matrix $S\in \RR^{r \times r}$. Hence, we can write the
evolution equation for the $K$ step~\eqref{eq:Kstep} in matrix formulation as
follows 
\begin{equation} \label{eq:Kstepmat}
  \partial_tK(t) = - \sum_{i=1}^d \overline{\partial}_{x_i} K(t)C_{1,v_i}^{\sf T} +
  \sum_{i=1}^d \text{diag}(E_{x_i}(K(t))) K(t) C_{2,v_i}^{\sf T} {\color{black},}
\end{equation}
where
\begin{equation*}
  \begin{aligned}
    K(t) = X(t)S(t), \quad K(t) \in \RR^{N_x\times r},  \\
    C_{1,v_i} =  \tilde{V}^{\sf T}\diag(\omega_{1,v_i})\tilde{V} \in \RR^{r \times r},  \\
    C_{2,v_i} =  \tilde{V}^{\sf T}\diag(\omega_{2,v_i})\overline{\partial}_{v_i}\tilde{V} \in \RR^{r \times r}{\color{black},}
  \end{aligned} 
\end{equation*}
{and \color{black}$\omega_{1,v_i}$ and $\omega_{2,v_i}$ are suitable quadrature weights}.
The $i$th component of the electric field has been denoted by 
$E_{x_i}(K(t)) \in \RR^{N_x}$. In addition, we have used $\overline{\partial}_{x_i}$
to denote the discretization of the spatial derivative operator. While this
operator can be represented as a matrix, in many cases it is more efficient to
directly compute its application to $K(t)$ (e.g.~in a stencil code or by using
FFTs). We also note that in order to compute the coefficients $C_{1,v_i}$ and
$C_{2,v_i}$ it would (obviously) be very inefficient to form the diagonal
matrix. Instead, our framework {\color{black} \texttt{Ensign}} provides the
function \texttt{coeff} that takes the matrices as well as a vector of weights
as input and computes the corresponding quadrature
(see {\color{black}Section~\ref{sec:usage} for more details}). 

For the evolution equation of the $S$ step~\eqref{eq:Sstep} we obtain 
\begin{equation} \label{eq:Sstepmat}
  \dot{S}(t) = \sum_{i=1}^d  D_{2,x_i} S(t) C_{1,v_i}^{\sf T}
                    -\sum_{i=1}^d D_{1,x_i}[E(S(t))] S(t) C_{2,v_i}^{\sf T}{\color{black},}
\end{equation}
where
\begin{equation*}
  \begin{aligned}
    D_{1,x_i}[E(S(t))] &=  \breve{X}^{\sf T}\diag(\omega_{1,x_i}^E)\breve{X} \in \RR^{r\times r},  \\
    D_{2,x_i} &=  \breve{X}^{\sf T}\diag(\omega_{2,x_i})\overline{\partial}_{x_i}\breve{X} \in \RR^{r\times r},
  \end{aligned} 
\end{equation*}
{\color{black} and again $\omega_{1,x_i}^E$ and $\omega_{2,x_i}$ are suitable
quadrature weights.}

Finally, for the evolution equation of the $L$ step~\eqref{eq:Lstep} we have
\begin{equation} \label{eq:Lstepmat}
  \partial_tL(t) = \sum_{i=1}^d \overline{\partial}_{v_i} L(t) D_{1,x_i}^{\sf T}
                   -\sum_{i=1}^d \text{diag}(\overline{v_i}) L(t) D_{2,x_i}^{\sf T}{\color{black},}
\end{equation}
where
\begin{equation*}
  L(t) = V(t)S(t)^{\sf T}, \quad L(t) \in \RR^{N_v\times r}{\color{black},}
\end{equation*}
and $\overline{v_i} \in \RR^{N_v}$ is the vector with the positions of the grid
points in velocity space.

{\color{black}In matrix formulation,} the equations for the electric field are given by 
\begin{equation*}
  \begin{aligned}
    &(E_{x_1}(f)(t),\ldots,E_{x_d}(f)(t)) = -\overline{\nabla}_x \Phi(f)(t),
    \quad E_{x_i}(f)(t) \in \RR^{N_x},\quad \Phi(f)(t) \in \RR^{N_x}{\color{black},} \\
    &-\overline{\Delta} \Phi(f)(t) = P(f)(t) + 1, \qquad P(f)(t) \in \RR^{N_x}{\color{black},}
  \end{aligned}
\end{equation*}
where the discretized charge density $P(f)(t)$ can be computed in terms of the
low-rank factors, depending on the need, as 
\begin{align}
  &P(K)(t) = - K(t)\tilde{P}, \qquad \tilde{P} = \tilde{V}^{\sf T}\omega_{v} \in \RR^r,
  \quad \omega_{v} \in \RR^{N_v}{\color{black},}  \label{eq:efK}\\
  &P(S)(t) = - \breve{X}S(t)\tilde{P}, \nonumber \\
  &P(L)(t) = - \breve{X}\overline{P}(L(t)),
  \qquad \overline{P}(L(t)) = L(t)^{\sf T}\omega_{v} \in \RR^r{\color{black}.} \label{eq:efL}
\end{align}
{\color{black} Here $\omega_{v}$ is a vector which collects suitable quadrature weights.}

Let us also note that the approximation of the {\color{black}particle-density}
function can be recovered at any moment from the low-rank factors by computing
\begin{equation*}
  F(t) = X(t)S(t)V(t)^{\sf T} \in \RR^{N_x\times N_v},
\end{equation*}
but clearly this is not needed for the low-rank projector-splitting algorithm,
and doing so would be extremely costly.

\subsection{Order 1 low-rank projector-splitting algorithm} \label{sec:foalgo}
The subflows corresponding to the $K$ step,
the $S$ step, and the $L$ step are then combined by a splitting scheme in order
to recover an approximation of the {\color{black}particle-density} function.
For a first order method it is clearly sufficient to consider a
Lie--Trotter splitting algorithm. A detailed description of the resulting
scheme is given in Algorithm~\ref{alg:firstorder}.
\begin{algorithm}
  \SetAlgoLined
  \KwIn{$X^0$, $S^0$, $V^0$ such that $f(0,x,v) \approx \sum_{i,j}X^0_i(x)S^0_{ij}V^0_j(v)$, {\color{black} time step} size $\tau$}
  \KwOut{$X^1$, $S^3$, $V^1$ such that $f(\tau,x,v) \approx \sum_{i,j}X^1_i(x)S^3_{ij}V^1_j(v)$}
  Compute $C_{1,v_i}$ and $C_{2,v_i}$, $i=1,\ldots,d$, using $V^0$\;
  Compute $K^0$ using $X^0$ and $S^0$\;
  Compute the electric field $E^0$ with {\color{black}equation}~\eqref{eq:efK} using $K^0$ and $V^0$\;
  Solve {\color{black}equation}~\eqref{eq:Kstepmat} with initial value $K^0$ and $E^0$ up to time $\tau$ to obtain $K^1$\;
  Perform a QR decomposition of $K^1$ to obtain $X^1$ and $S^1$\;
  Compute $D_{1,x_i}$ and $D_{2,x_i}$, $i=1,\ldots,d$, using $E^0$ and $X^1$\;
  Solve {\color{black}equation}~\eqref{eq:Sstepmat} with initial value $S^1$ and $E^0$ up to time $\tau$ to obtain $S^2$\;
  Compute $L^0$ using $V^0$ and $S^2$\;
  Solve {\color{black}equation}~\eqref{eq:Lstepmat} with initial value $L^0$ and $E^0$ up to time $\tau$ to obtain $L^1$\;
  Perform a QR decomposition of $L^1$ to obtain $V^1$ and $S^3$\;
  \caption{First order {\color{black}dynamical low-rank} integrator for {\color{black} the Vlasov--Poisson equations}~\eqref{eq:vp}.}
  \label{alg:firstorder}
\end{algorithm}

Remark that the computation of the electric field is performed only once at
the beginning of the time step. This is not a restriction in the context of a
Lie--Trotter splitting, as fixing the electric field at each time step still
results in a first order approximation.
In principle, any numerical method can be employed to integrate in time
{\color{black} equations}~\eqref{eq:Kstepmat}--\eqref{eq:Lstepmat}.
We will discuss our proposal, based on the choice of a spectral phase space
discretization, in Section~\ref{sec:tintKSL}. 
Moreover, after performing the $K$ and the $L$
steps, in order to proceed with the algorithm we need to recover the
orthonormal functions $X_i$, $V_j$ and the coupling coefficients $S_{ij}$.
This can be accomplished, for example, by a QR or an SVD decomposition.
{\color{black}Finally, we note that overall storage and computational costs of the dynamical
low-rank algorithm, for $n_{x_i}=n_{v_i}=n$, scale as $\OO(rn^d)$ and $\OO(r^2n^d)$, respectively.
For more details we refer the reader to~\cite{einkemmer2018low}.}

\subsection{Order 2 low-rank projector-splitting algorithm} \label{sec:soalgo}
A straightforward generalization to a second order integrator by employing a
Strang splitting procedure instead of a Lie--Trotter one is not possible.
Indeed, if we freeze the electric field at the beginning of the time step,
we still end up with a first order algorithm. To overcome this problem, an
almost symmetric Strang splitting scheme is proposed in~\cite{einkemmer2018low};
however, in order to achieve full second order, that algorithm requires several
updates of the electric field, which in turn translates into high computational
effort. We propose here a slightly different strategy to obtain a second order
scheme, listed in detail in Algorithm~\ref{alg:secondorder}. The underlying
idea is that we compute an approximation of the electric field at time
$\tau/2$ of (local) second order by means of Algorithm~\ref{alg:firstorder}.
Then, we restart the integration with a classic Strang splitting scheme
employing as constant electric field the approximation at the half step.
Mathematically, this can {\color{black} still} be analyzed as an almost symmetric
splitting scheme (see~\cite{einkemmer2014almost1,einkemmer2014almost2}),
{\color{black}and as for Algorithm~\ref{alg:firstorder} the storage and
computational costs scale as $\OO(rn^d)$ and $\OO(r^2n^d)$, respectively.}

\begin{algorithm}
  \SetAlgoLined
  \KwIn{$X^0$, $S^0$, $V^0$ such that $f(0,x,v) \approx \sum_{i,j}X^0_i(x)S^0_{ij}V^0_j(v)$, {\color{black} time step} size $\tau$}
  \KwOut{$X^3$, $S^7$, $V^1$ such that $f(\tau,x,v) \approx \sum_{i,j}X^3_i(x)S^7_{ij}V^1_j(v)$}
  {\color{black} Perform steps 1--9 of Algorithm~\ref{alg:firstorder} with time step size $\tau/2$\;}
    Compute the electric field $E^{1/2}$ with {\color{black} equation}~\eqref{eq:efL} using $X^1$ and $L^1$ {\color{black}from step 1}.\;
  Solve {\color{black} equation}~\eqref{eq:Kstepmat} with initial value $K^0$ and $E^{1/2}$ up to time $\tau/2$ to obtain $K^2$\;
  Perform a QR decomposition of $K^2$ to obtain $X^2$ and $S^3$\;
  Compute $D_{1,x_i}$ and $D_{2,x_i}$, $i=1,\ldots,d$, using $E^{1/2}$ and $X^2$\;
  Solve {\color{black} equation}~\eqref{eq:Sstepmat} with initial value $S^3$ and $E^{1/2}$ up to time $\tau/2$ to obtain $S^4$\;
  Compute $L^1$ using $V^0$ and $S^4$\;
  Solve {\color{black} equation}~\eqref{eq:Lstepmat} with initial value $L^1$ and $E^{1/2}$ up to time $\tau$ to obtain $L^2$\;
  Perform a QR decomposition of $L^2$ to obtain $V^1$ and $S^5$\;
  Compute $C_{1,v_i}$ and $C_{2,v_i}$, $i=1,\ldots,d$, using $V^1$\;
  Solve {\color{black} equation}~\eqref{eq:Sstepmat} with initial value $S^5$ and $E^{1/2}$ up to time $\tau/2$ to obtain $S^6$\;
  Compute $K^3$ using $X^2$ and $S^6$\;
  Solve {\color{black} equation}~\eqref{eq:Kstepmat} with initial value $K^3$ and $E^{1/2}$ up to time $\tau/2$ to obtain $K^4$\;
  Perform a QR decomposition of $K^4$ to obtain $X^3$ and $S^7$\;

  \caption{Second order {\color{black}dynamical low-rank} integrator for {\color{black} the Vlasov--Poisson equations}~\eqref{eq:vp}.}
  \label{alg:secondorder}
\end{algorithm} 

\section{Time and space discretization of $K$, $S$ and $L$ steps} \label{sec:tintKSL}
As already mentioned in Section~\ref{sec:matform}, in principle any numerical
scheme can be used to integrate in time equations~\eqref{eq:Kstepmat}--\eqref{eq:Lstepmat}.
However, depending on the selected phase-space discretization, some choices
can be more adequate than others. In particular, we describe here in detail an
exponential integrator based strategy which uses a Fourier spectral discretization
for both space and velocity
variables. The method converges rapidly in space and velocity (owing to the
spectral discretization) and despite being fully explicit does not suffer from
a CFL induced {\color{black} step size} restriction in time.

Let us consider the $K$ step~\eqref{eq:Kstepmat}.
{\color{black} As we are interested in performing 6D simulations, }
for clarity of exposition we will only consider the case {\color{black} $d=3$} here.
{\color{black} The (simpler) cases $d=1$ and $d=2$ can be treated similarly.}
{\color{black} Then,} by performing a further splitting of the $K$ step we obtain
the following three equations
\begin{align}
  &\partial_tK_1(t) = -\overline{\partial}_{x_1}K_1(t)C_{1,v_1}^{\sf T}, \label{eq:k1split}\\
  &\partial_tK_2(t) = -\overline{\partial}_{x_2}K_2(t)C_{1,v_2}^{\sf T},\label{eq:k2split}\\
  &\partial_tK_3(t) =-\overline{\partial}_{x_3}K_3(t)C_{1,v_3}^{\sf T}
  + \sum_{i=1}^3 \text{diag}(E_{x_i}) K_3(t) C_{2,v_i}^{\sf T}.\label{eq:k3split} 
\end{align}
{\color{black}Note that, in principle, the summation term related to the electric
field could be put in any of the three equations \eqref{eq:k1split}--\eqref{eq:k3split},
without any substantial change.
On the other hand, in a general setting, there would not be advantages in treating
this term in a separate flow, because there is no exact solution in Fourier space and more
splitting error would be generated.}

Applying then a Fourier transform in the $x_i$ variables (denoted by $\mathcal{F}_{x_i}$)
we obtain
\begin{align*}
  &\partial_t\hat{K}_1(t) = -D^\mathcal{F}_{x_1}\hat{K}_1(t)C_{1,v_1}^{\sf T},  \\
  &\partial_t\hat{K}_2(t) = -D^\mathcal{F}_{x_2}\hat{K}_2(t)C_{1,v_2}^{\sf T},  \\
  &\partial_t\hat{K}_3(t) =-D^\mathcal{F}_{x_3}\hat{K}_3(t)C_{1,v_3}^{\sf T} 
  +  \sum_{i=1}^3 \mathcal{F}_{x_3}(\text{diag}(E_{x_i}) K_3(t))  C_{2,v_i}^{\sf T},
\end{align*}
where {\color{black} $\hat{K}_i(t) = \mathcal{F}_{x_i}(K_i(t))$} and
$D^{\mathcal{F}}_{x_i}$ is a diagonal matrix containing the coefficients
stemming from the differential operator $\overline{\partial}_{x_i}$ in Fourier
space. Then, as the matrices $C^{\sf T}_{1,v_i}$ are symmetric by construction,
it is possible to diagonalize them so that $C_{1,v_i}^{\sf T} = T_{v_i}D_{v_i}T_{v_i}^{\sf T}$.
We note that this operation is computationally cheap as the matrices involved
have only size $r\times r$.
By performing the substitution $\hat{M}_i(t)=\hat{K}_i(t)T_{v_i}$ we have
\begin{align}
  &\partial_t\hat{M}_1(t) = -D^\mathcal{F}_{x_1}\hat{M}_1(t)D_{v_1}{\color{black},} \label{eq:f11split} \\
  &\partial_t\hat{M}_2(t) = -D^\mathcal{F}_{x_2}\hat{M}_2(t)D_{v_2}{\color{black},} \label{eq:f21split} \\
  &\partial_t\hat{M}_3(t) =-D^\mathcal{F}_{x_3}\hat{M}_3(t)D_{v_3} 
  +\sum_{i=1}^3 \mathcal{F}_{x_3}(\text{diag}(E_i) K_3(t))  C_{2,v_i}^{\sf T} T_{v_3}{\color{black}.} \label{eq:s1split}
\end{align}
At this point, equations~\eqref{eq:f11split} and~\eqref{eq:f21split} can be
solved exactly in time by means of independent pointwise operations, while
{\color{black} equation}~\eqref{eq:s1split} can be solved efficiently by means
of a first or second order exponential Runge--Kutta method (see~\cite{hochbruck2010}
for a survey), again just using independent pointwise operations. We choose to
use exponential integrators in the time evolution of this step because in this
way we remove any CFL-like restriction of the {\color{black} step size} coming
from the stiffness of the spatial derivative. Moreover, as everything is
written in terms of independent pointwise operations, the computation of the
single {\color{black} flow} can be performed completely in parallel.

Finally, coming back to the original variables $\hat{K}_i$ and performing an
inverse Fourier transform, we obtain approximate solutions for the
{\color{black} equations involved}. Embedding this procedure in a splitting context returns
the desired approximation of the evolution equation for the $K$ step. In
particular, for the first order method described in Algorithm~\ref{alg:firstorder}
it is enough to perform a Lie--Trotter splitting, while a (classical) Strang splitting
procedure is needed for the second order method presented in
Algorithm~\ref{alg:secondorder}.

Concerning the integration of the evolution equation of the
$L$ step~\eqref{eq:Lstepmat}, similar considerations apply. 
Finally, concerning the $S$ step~\eqref{eq:Sstepmat}, it is an $r\times r$ problem
and there is no source of stiffness in it. Hence, we perform its time integration
by means of the classical explicit fourth order Runge--Kutta scheme RK4.
{\color{black}In principle it would be sufficient to use a first order scheme,
in the context of Algorithm~\ref{alg:firstorder}, and a second order scheme,
in the context of~Algorithm \ref{alg:secondorder}. However, since the $S$ step
is cheap, we can use a higher order approximation at negligible additional
computational cost and reduce the error term associated with integrating $S$
from the numerical scheme.}

\section{The \texttt{Ensign} framework and implementation} \label{sec:usage}
{\color{black} As presented in Sections~\ref{sec:lrcont}--\ref{sec:tintKSL},}
in the context of dynamical low-rank algorithms a function $f(t,x,v)$ 
is approximated as $f(t,x,v)\approx\sum_{i,j}X_i(t,x)S_{ij}(t)V_j(t,v)$,
where the indexes $i$ and $j$ run from $1$ to $r$ and $r$ is the
chosen approximation rank. The quantities $X_i(t,x)$, $S_{ij}(t)$ and $V_j(t,v)$
constitute the so called low-rank factors and, after discretization in $x$ and $v$,
they can be expressed as matrices{\color{black}, allowing us to  write the scheme
in matrix formulation}.
Therefore, independently of the specific case under consideration, the common
key point for an efficient implementation of dynamical low-rank algorithms for
kinetic equations, both on CPU and GPU based systems, is the fast computation
of {\color{black}operations on matrices and vectors (e.g. matrix-matrix and matrix-vector products, certain pointwise operations, etc)}. For every modern computer
architecture, we have at our disposal heavily optimized routines to perform
such operations, usually referred to as BLAS. For example, Intel MKL~\cite{mkl}
and OpenBLAS~\cite{xianyi2012model} are available for CPU based systems while we
have cuBLAS~\cite{cublas} and MAGMA~\cite{ntd10_vecpar} for NVIDIA GPUs. Among
their many features, all these libraries are equipped with multithreaded versions of BLAS routines.
{\color{black}When possible we use these libraries heavily. However, we note that
there are certain operations that we need to perform in the context of a low-rank 
algorithm that are not part of BLAS.}

The main idea behind \texttt{Ensign} is to provide the user a collection of
structures and functions in order to compute and manipulate \textit{easily}
the arising quantities in dynamical low-rank algorithms.
{\color{black}Let us note that our goal here is to provide primitives that allow
the user to implement dynamical low-rank approximations on a high-level.
Thus, we are concerned with relevant data structures for dynamical low-rank algorithms,
for computing the coefficients that appear in the approximations (which can
have, in general, two or more indices and also spatial dependences), for
performing certain operations such as initialization, addition or truncation
on low-rank approximations, orthogonalization with respect to arbitrary inner products,
and writing such low-rank approximations to disk.
This could then be complemented by the user of \texttt{Ensign} with libraries that
perform spatial and temporal discretization.}
Our framework is written in C++ programming language and uses CUDA internally for
the GPU code.

We now illustrate some of its features with {\color{black} the aid of the first 
order projector-splitting dynamical low-rank algorithm presented in Section~\ref{sec:foalgo},
assuming an underlying six-dimensional phase space.
To this aim, in Figure~\ref{code:dlr} we translate in source code some lines of
the pseudocode of Algorithm~\ref{alg:firstorder}, and in Figure~\ref{code:dlrinit}
we show how it is possible to initialize the low-rank factors.
We can immediately note how the quantities arising from the mathematical equations can be naturally}
translated into the structures and functions provided by our framework.
\begin{figure}[!htb]
\begin{lstlisting}
// Declaration and initialization
/* See Figure 2 */
gram_schmidt gs(&blas);
multi_array<double,1> w1v1(stloc::host); // stloc::device if on GPU
multi_array<double,2> C1v1(stloc::host);
/* ... */
// Line 1: Compute C coefficients
coeff(lr_st.V, lr_st.V, w1v1, C1v1, blas);
coeff(lr_st.V, lr_st.V, w1v2, C1v2, blas);
coeff(lr_st.V, lr_st.V, w1v3, C1v3, blas);
coeff(lr_st.V, dV0_v1, w2v1, C2v1, blas);
coeff(lr_st.V, dV0_v2, w2v2, C2v2, blas);
coeff(lr_st.V, dV0_v3, w2v3, C2v3, blas);
// Line 2: Compute K0
tmpX = lr_st.X;
blas.matmul(tmpX, lr_st.S, lr_st.X);
// Line 3: Compute electric field
/* ... */
// Line 4: Solve K step
/* ... */
// Line 5: Perform QR decomposition
gs(lr_st.X, lr_st.S, ip_xx);
/* ... */
// Line 6: Compute D coefficients
coeff(lr_st.X, lr_st.X, wE1x1, D1x1, blas);
coeff(lr_st.X, lr_st.X, wE1x2, D1x2, blas);
coeff(lr_st.X, lr_st.X, wE1x3, D1x3, blas);
coeff(lr_st.X, dX1_x1, w2x1, D2x1, blas);
coeff(lr_st.X, dX1_x2, w2x2, D2x2, blas);
coeff(lr_st.X, dX1_x3, w2x3, D2x3, blas);
// Line 7: Solve S step
/* ... */
// Line 8: Compute L0
tmpV = lr_st.V;
blas.matmul_transb(tmpV, lr_st.S, lr_st.V);
// Line 9: Solve L step
/* ... */
// Line 10: Perform QR decomposition
gs(lr_st.V, lr_st.S, ip_vv);
transpose_inplace(lr_st.S);
\end{lstlisting}
\caption{{\color{black}Sketch of a} C++ implementation of
Algorithm~\ref{alg:firstorder} using the \texttt{Ensign} framework.
{\color{black}To perform computation on the GPU, it is enough to use \texttt{d\_lr\_st}
instead of \texttt{lr\_st}, and to declare the relevant \texttt{multi\_array}s with
\texttt{stloc::device}. The syntax {\color{black}\texttt{/*...*/}} indicates code not reported for simplicity
of exposition.}}\label{code:dlr}
\end{figure}
In particular, in terms of data storage we employ
a structure \texttt{multi\_array} which lets us easily define vectors
(\verb+w1v1+, for example) and matrices (\verb+C1v1+, for example) both in CPU and
in GPU memory, depending on the need. This structure is 
also enriched with some user-friendly functions and operators which are useful 
to perform basic operations between \texttt{multi\_array}s, such as sum,
difference, multiplication with a scalar, and to transfer data from/to
CPU/GPU memory (the latter can be simply done by assignment, as commented in
Figure~\ref{code:dlrinit}).
{\color{black}For convenience of the user, we provide also a structure \texttt{lr2},
which contains three 2D \verb+multi_array+s (\verb+X+, \verb+S+ and \verb+V+)
that}
reflect the evaluation of the low-rank factors on a discretized
 grid {\color{black}and their coupling coefficients}. In particular, the degrees of freedom are linearized so that
each column corresponds to the discretized version of a single low-rank factor.
While the specific example of Figure~\ref{code:dlrinit} is presented in the context of a
uniform space discretization, any other kind of discretization (with an arbitrary ordering
of the nodes) would work in a straightforward way as well.
\begin{figure}[!htb]
\begin{lstlisting}
typedef ptrdiff_t Index;
array<Index,3> N_xx, N_vv;
array<double,3> lim_xx, lim_vv, h_xx, h_vv;
Index r;
blas_ops blas;
vector<const double*> X0, V0;

// Initialize N_xx, lim_xx, h_xx, N_vv, lim_vv, h_vv and r on CPU
/* ... */

Index N_xx_m = N_xx[0]*N_xx[1]*N_xx[2];
Index N_xx_m = N_vv[0]*N_vv[1]*N_vv[2];

// Define inner products that are used in the algorithm
auto ip_xx = inner_product_from_const_weight(h_xx[0]*h_xx[1]*h_xx[2], N_xx_m);
auto ip_vv = inner_product_from_const_weight(h_vv[0]*h_vv[1]*h_vv[2], N_vv_m);

// Initialize the low-rank structure for the initial value
// (the initial value is given by X0 and V0 and can usually
// be easily determined from the problem)
// To illustrate data movement (see below) we perform the
// initialization on the CPU and then transfer the result
// to the GPU.

lr2<double> lr_st(r,{N_xx_m,N_vv_m});

// Set up X0 and V0 
/* ... */

initialize(lr_st, X0, V0, ip_xx, ip_vv, blas);

// Assignment of two lr2 or multi_arrays copies from CPU to GPU
lr2<double> d_lr_st(r,{N_xx_m,N_vv_m},stloc::device); // on GPU
d_lr_st = lr_st;
\end{lstlisting}
\caption{{\color{black}Sketch of a} C++ implementation for the
initialization of low-rank factors using the \texttt{Ensign} framework.
{\color{black} \texttt{N\_xx},
\texttt{lim\_xx} and \texttt{h\_xx} are arrays which contain the number of
discretization points, the left extremes of the space domain and the grid spacing 
for each direction, respectively (similarly for \texttt{N\_vv},
\texttt{lim\_vv} and \texttt{h\_vv}, which are related to the velocity domain).
\texttt{r} is the approximation rank, while
\texttt{lr\_st} and \texttt{d\_lr\_st} are structures which contain 
2D \texttt{multi\_arrays} that reside on the CPU and on the GPU, respectively.
The syntax {\color{black}\texttt{/*...*/}} indicates code not reported for simplicity
of exposition.}
}\label{code:dlrinit}
\end{figure}

We use C++ templates in order to abstract the underlying architecture on which 
the code is run. 
Depending on whether a function is called with 
arguments that reside on the CPU or on the GPU, an efficient implementation 
suitable for that hardware architecture is selected.
Indeed, whether the code runs on the CPU or on the GPU is never explicitly 
specified in the code example in Figure~\ref{code:dlr}, because it is automatically detected from 
the storage location of the input
arguments of the functions. Thus, the algorithmic part of the implementation 
is completely independent of what computer hardware the code is eventually run on.

Concerning BLAS operations, \texttt{Ensign} provides the structure \verb+blas_ops+
which contains wrappers for matrix multiplications (e.g. \verb+matmul+ 
and \verb+matmul_transb+) and handles needed to call properly these routines on the GPU.
Then, our framework is equipped with a
function to compute the $C$ and $D$ integral coefficients, namely \texttt{coeff}.
Depending on the input matrices and the vector of weights, 
the function computes the corresponding matrix of coefficients. 
As an example, given the 1D quadrature weight \texttt{multi\_array} \texttt{w1v1} 
and the 2D \texttt{multi\_array} \texttt{lr\_st.V}, the coefficient matrix $C_{1,v_1}$
{\color{black}can be} computed {\color{black} using the command} in line~8 of
{\color{black} the code in} Figure~\ref{code:dlr}. 
Also, we can find in our framework the structure \verb+gram_schmidt+, which
contains a function to compute the QR decomposition {\color{black}of a matrix with 
a generic inner product}.
It is based on a modified {\color{black}Gram--Schmidt} algorithm written as much as possible in matrix formulation, so that again the internal computations are automatically performed in parallel by means of calls to appropriate BLAS. {\color{black}Modified Gram-Schmidt is used here because it is easy to parallelize and can operate purely in terms of inner products, even if the associated degrees of freedom are not stored directly as a vector. The latter is important in the case of hierarchical low-rank approximations such as the those considered in \cite{einkemmer2018low}.}
For example, given the 2D \texttt{multi\_array}s 
\texttt{lr\_st.X} and \texttt{lr\_st.S} and {\color{black} the
inner product \texttt{ip\_xx}}, the call to perform 
the QR decomposition of \verb+lr_st.X+ is given in line~22 of Figure~\ref{code:dlr}.

Finally, we want to emphasize that while we have illustrated our software framework for
a simple projector-splitting based dynamical low-rank integrator, every effort has been 
made in designing \texttt{Ensign} to allow also the implementation of other 
dynamical low-rank integrators, such as the recently proposed unconventional 
integrator~\cite{ceruti2021unconventional} or the conservative dynamical 
low-rank integrator~\cite{einkemmer2021mass}. 
Moreover, we again point out that the user is completely free to choose any space or time discretization 
appropriate to the problem. The only requirement in terms of space discretization 
is that the degrees of freedom of the low-rank factors have to be collected in suitable 
matrices by means of an index linearization. 
This is certainly possible for all the commonly used space discretization strategies.

\section{Numerical experiments \label{sec:numerics}}

In this section we will present some numerical results and validate the
implementations of the algorithms described in Section~\ref{sec:matform}.
The developed code solves the 6D Vlasov--Poisson {\color{black}equations}~\eqref{eq:vp}
and uses the framework \texttt{Ensign}. All the experiments in
this section have been performed {\color{black}in double precision arithmetic} with the aid of a single NVIDIA Tesla A100 card
(theoretical peak memory bandwidth of 1555 GB/s and peak floating point processing 
power for double precision of 9.7 TFlops), equipped with 40 GB of RAM. 

\subsection{Orders of convergence}\label{sec:ordconv}
First of all, we check the implementation of the low-rank projector-splitting
algorithms by computing numerically the order of convergence of the methods.
To this purpose, we consider a 6D linear Landau problem posed on the domain
$\Omega = (0,4\pi)^3\times(-6,6)^3$ with an initial particle-density given
by
\begin{equation}\label{eq:ll}
f_0(x_1,x_2,x_3,v_1,v_2,v_3) = \frac{1}{\sqrt{(2\pi)^3}}
\ee^{-(v_1^2+v_2^2+v_3^2)/2}(1+\alpha_1\cos(\kappa_1 x_1)+\alpha_2\cos(\kappa_2 x_2)+\alpha_3\cos(\kappa_3 x_3)).
\end{equation}
The parameters are set to $\alpha_1 = \alpha_2 = \alpha_3 = 10^{-2}$ and 
$\kappa_1 = \kappa_2 = \kappa_3 = \frac{1}{2}$. We consider the problem with 
periodic boundary conditions in all directions and integrate it up to final 
time $T=1$. Concerning the space discretization, we take $32^3$ 
as number of discretization points for both space and velocity variables.
The rank of the solution is fixed to $r=10$.
As a reference solution, we take the result of the second order low-rank projector-splitting
algorithm with $m=2000$ {\color{black} time steps} (time step size $\tau = 5\cdot 10^{-4}$).

We also consider a 6D two stream instability {\color{black}problem} defined on
the domain $\Omega = (0,10\pi)^3\times(-9,9)^3$ with initial distribution
\begin{equation}\label{eq:ts}
\begin{aligned}
f_0(x_1,x_2,x_3,v_1,v_2,v_3) = &\frac{1}{\sqrt{(8\pi)^3}}\left(\ee^{-(v_1-\overline{v}_1)^2/2}+\ee^{-(v_1-\tilde{v}_1)^2/2}\right) \\
&\times\left(\ee^{-(v_2-\overline{v}_2)^2/2}+\ee^{-(v_2-\tilde{v}_2)^2/2}\right) \\
&\times\left(\ee^{-(v_3-\overline{v}_3)^2/2}+\ee^{-(v_3-\tilde{v}_3)^2/2}\right) \\
&\times(1+\alpha_1\cos(\kappa_1 x_1)+\alpha_2\cos(\kappa_2 x_2)+\alpha_3\cos(\kappa_3 x_3)).
\end{aligned}
\end{equation}
In this case the parameters are given by
$\alpha_1 = \alpha_2 = \alpha_3 = 10^{-3}$, $\kappa_1 = \kappa_2 = \kappa_3 = \frac{1}{5}$,
$\overline{v}_1 = \frac{5}{2}$, $\overline{v}_2 = \overline{v}_3 = 0$,
$\tilde{v}_1 = -\frac{5}{2}$, $\tilde{v}_2 = -\frac{9}{4}$ and $\tilde{v}_3 = -2$.
As for linear Landau damping, the problem is equipped with periodic boundary conditions in 
all directions and the rank is fixed to $r=10$. We perform simulations up to final
time $T = \frac{1}{20}$ with $32^3$ discretization points for both space and velocity
variables. We again consider
as a reference solution the results of the second order algorithm with $m=2000$ 
{\color{black} time steps} ({\color{black} time step} size $\tau = 2.5 \cdot 10^{-5}$).

\begin{figure}[!htb]
\centering
%
%
\begin{tikzpicture}

\begin{axis}[%
width=2.13in,
height=2.13in,
at={(0.944in,0.642in)},
scale only axis,
xmode=log,
xmin=39.8,
xmax=80,
xlabel style={font=\color{white!15!black}},
xlabel={$m$},
xminorticks=true,
xticklabels={40,50,60,70,80},xtick={40,50,60,70,80},
ymode=log,
ymin=1e-07,
ymax=2e-04,
yminorticks=true,
ylabel style={font=\color{white!15!black}},
ylabel={Error},
axis background/.style={fill=white},
title={Linear Landau}
]
\addplot [color=black, dashed, forget plot]
  table[row sep=crcr]{%
40	0.00011784783489326\\
80	5.89239174466298e-05\\
};
\addplot [ultra thick, color=red, only marks, mark=x, mark options={solid, red, scale=1.5}, forget plot]
  table[row sep=crcr]{%
40	0.000116652111206601\\
50	9.37036242234429e-05\\
60	7.82990028718939e-05\\
70	6.72437845370483e-05\\
80	5.89239174466298e-05\\
};
\addplot [color=black, dashdotted, forget plot]
  table[row sep=crcr]{%
40	6.34155149506572e-07\\
80	1.58538787376643e-07\\
};
\addplot [ultra thick, color=green, only marks, mark=o, mark options={solid, green, scale = 1.5}, forget plot]
  table[row sep=crcr]{%
40	6.36066065354065e-07\\
50	4.06698385905703e-07\\
60	2.82215436703763e-07\\
70	2.0720276418421e-07\\
80	1.58538787376643e-07\\
};
\end{axis}

\begin{axis}[%
width=2.13in,
height=2.13in,
at={(4.141in,0.642in)},
scale only axis,
xmode=log,
xmin=39.8,
xmax=80,
xlabel style={font=\color{white!15!black}},
xlabel={$m$},
xminorticks=true,
xticklabels={40,50,60,70,80},xtick={40,50,60,70,80},
ymode=log,
ymin=3.76511166591576e-10,
ymax=3.54101990275081e-08,
yminorticks=true,
ylabel style={font=\color{white!15!black}},
ylabel={Error},
axis background/.style={fill=white},
title={Two stream instability}
]
\addplot [color=black, dashed, forget plot]
  table[row sep=crcr]{%
40	2.54101990275081e-08\\
80	1.2705099513754e-08\\
};
\addplot [ultra thick, color=red, only marks, mark=x, mark options={solid, red, scale=1.5}, forget plot]
  table[row sep=crcr]{%
40	2.5302479822387e-08\\
50	2.01922322338331e-08\\
60	1.68753374259117e-08\\
70	1.44966544213943e-08\\
80	1.2705099513754e-08\\
};
\addplot [color=black, dashdotted, forget plot]
  table[row sep=crcr]{%
40	1.50604466636631e-09\\
80	3.76511166591577e-10\\
};
\addplot [ultra thick, color=green, only marks, mark=o, mark options={solid, green, scale = 1.5}, forget plot]
  table[row sep=crcr]{%
40	1.483879677285e-09\\
50	9.59640979538395e-10\\
60	6.68386232193247e-10\\
70	4.91497138446623e-10\\
80	3.76511166591577e-10\\
};
\end{axis}

\end{tikzpicture}%
\caption{Orders of convergence for first order (red crosses) and second order (green circles)
low-rank  projector-splitting algorithms~{\color{black}\ref{alg:firstorder} and ~\ref{alg:secondorder}}. 
Left plot: linear Landau {\color{black}damping~\eqref{eq:ll} with $T=1$}. Right plot: two stream
instability {\color{black}problem~\eqref{eq:ts} with $T=\frac{1}{20}$}.
{\color{black}In both cases, the rank is set to $r=10$}.
The (relative) error is computed in maximum norm at the final time for a number of {\color{black} time steps}
equal to $m=40,50,60,70,80$, {\color{black}with respect to a reference solution
produced with Algorithm~\ref{alg:secondorder} and $m=2000$ time steps}.
The dashed and dashed-dotted lines are reference lines of slope -1 and -2, respectively.
\label{fig:order} }
\end{figure}
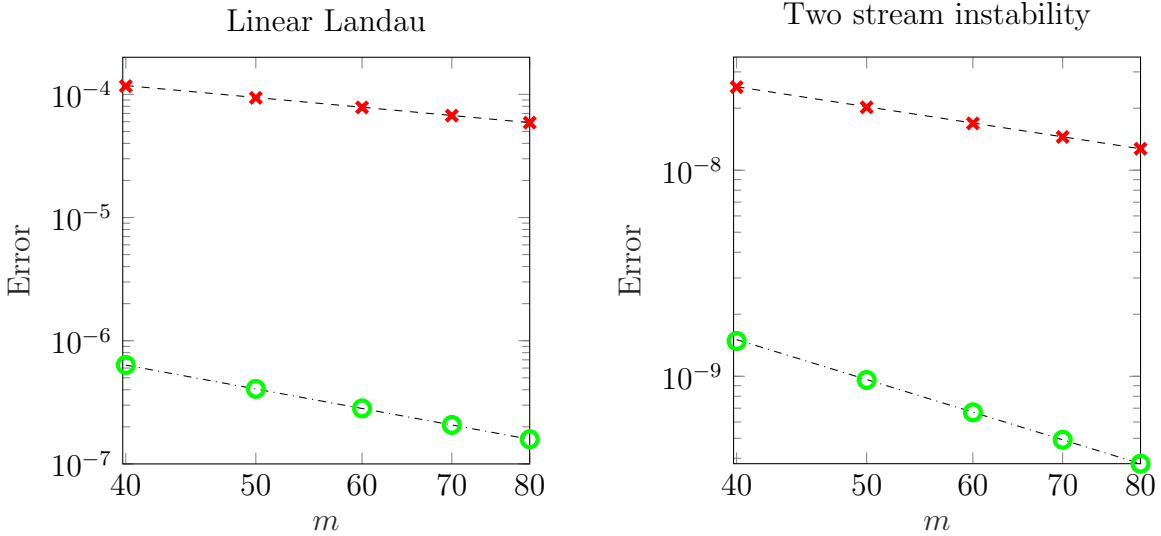

The results for both problems are collected in Figure~\ref{fig:order}. In 
each of the two cases, we can clearly see that the first and second order algorithms 
show the expected order of convergence.

\subsection{Linear Landau simulation} \label{sec:flagll}
We consider again the 6D linear Landau {\color{black}problem}~\eqref{eq:ll} with periodic boundary 
conditions and the same set of parameters. However, now we pick $64^3$ discretization
points for the space variables and $256^3$ for the velocity ones.
The approximation rank is fixed to $r=10$, {\color{black} and we perform two simulations
with time step sizes $\tau=10^{-1}$ and $\tau=10^{-2}$, respectively.}
We emphasize that a direct (Eulerian or semi-Lagrangian) Vlasov solver would
require {\color{black}a total number of degrees of freedom of approximately $4.4\cdot10^{12}$, and}
at least $70$ TB of main memory (RAM) to perform the simulations {\color{black}in double precision arithmetic}.
This would clearly only be feasible on a supercomputer. 
Our dynamical low-rank simulation, in contrast, runs on a single NVIDIA A100 equipped with 40 GB
of memory, and the number of degrees of freedom are $10\cdot64^3 + 10\cdot256^3 \approx 1.7 \cdot 10^{8}$. The computational time of execution is approximately 
{\color{black} six minutes for the simulation with $\tau=10^{-1}$} and an hour 
{\color{black} for the one with $\tau=10^{-2}$}.

\begin{figure}[!htb]
\centering
\input{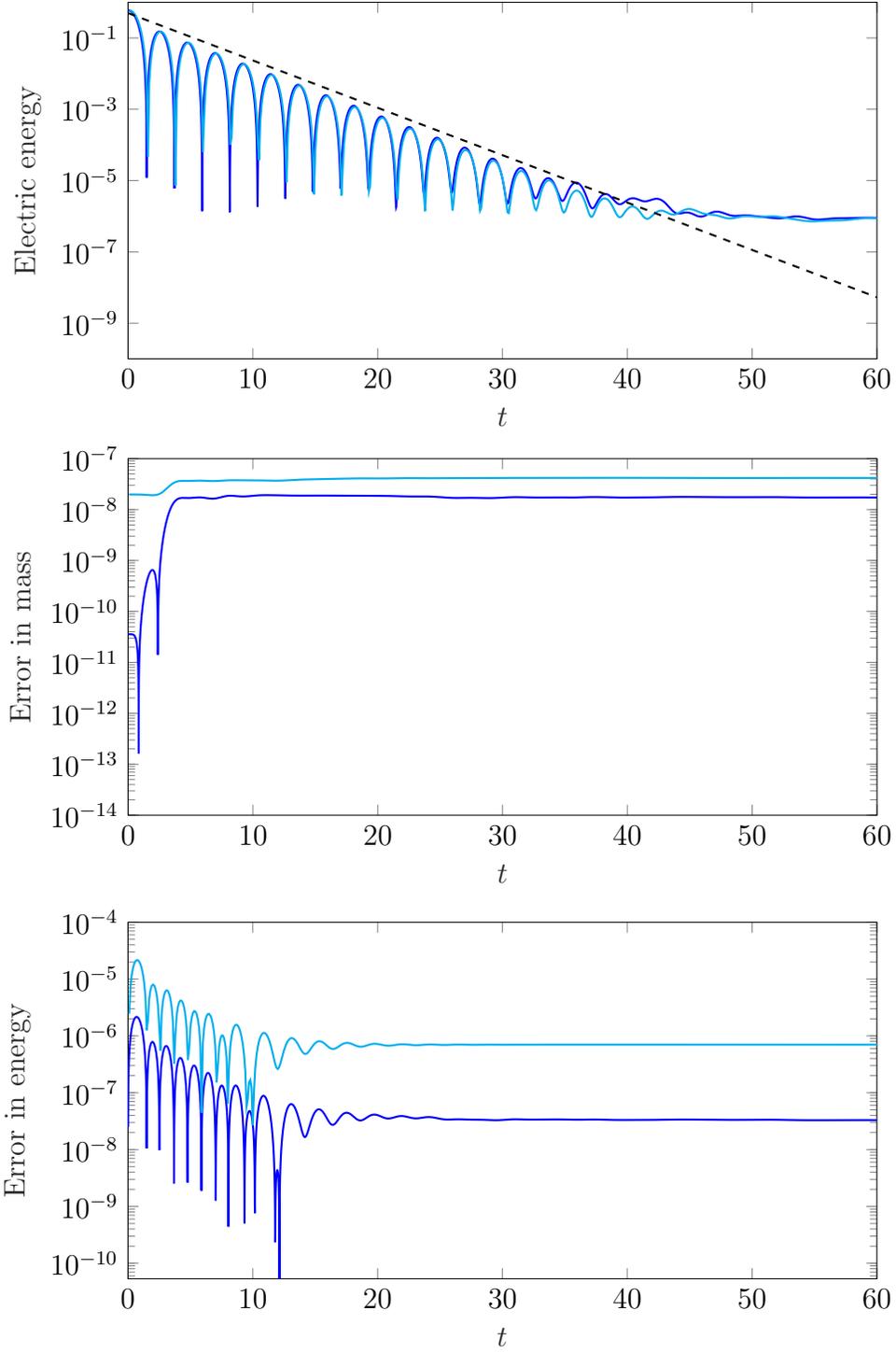}
\caption{Linear Landau simulation with $64^3$ space discretization points, $256^3$
velocity discretization points, 
final time $T=60$, rank $r=10$ and {\color{black} time step sizes $\tau=10^{-1}$ 
(cyan line) and $\tau=10^{-2}$ (blue line), see Section~\ref{sec:flagll}}.
The second order low-rank projector-splitting
algorithm~\ref{alg:secondorder} is employed. Top plot: behavior of electric energy, with reference decay rate.
Center plot: error in mass (relative). Bottom plot: error in total energy (relative).}
\label{fig:flagll}
\end{figure}

The results obtained, in terms of electric energy, error in mass and 
error in total energy, are summarized in Figure~\ref{fig:flagll}. We can 
observe that the electric energy shows the expected theoretical exponential rate 
of decay up to approximately $10^{-6}$, {\color{black}with very similar 
results for both the time step sizes. In this sense, what dominates after time $t=40$}
is the low-rank error, and we basically enter into a stagnation region
(see Section~\ref{sec:diffrank} and Figure~\ref{fig:diffrankll} for simulations with higher ranks).
Concerning the errors in mass and in total energy (both quantities are
conserved by the original {\color{black}equations}), even though the proposed
low-rank projector-splitting integrator does not preserve a priori any quantity
we obtain conservation {\color{black}of mass up to roughly $3\cdot10^{-8}$} 
and up to {\color{black}$7 \cdot 10^{-7}$ or $4 \cdot 10^{-8}$ for the energy,
depending on the time step size. In this sense, the simulation with smaller
time step size produces better results, meaning that for these quantities we 
still did not encounter a bound stemming from the low-rank truncation}.

\subsection{Two stream instability simulation}\label{sec:flagts}
\begin{figure}[!htb]
\centering
\input{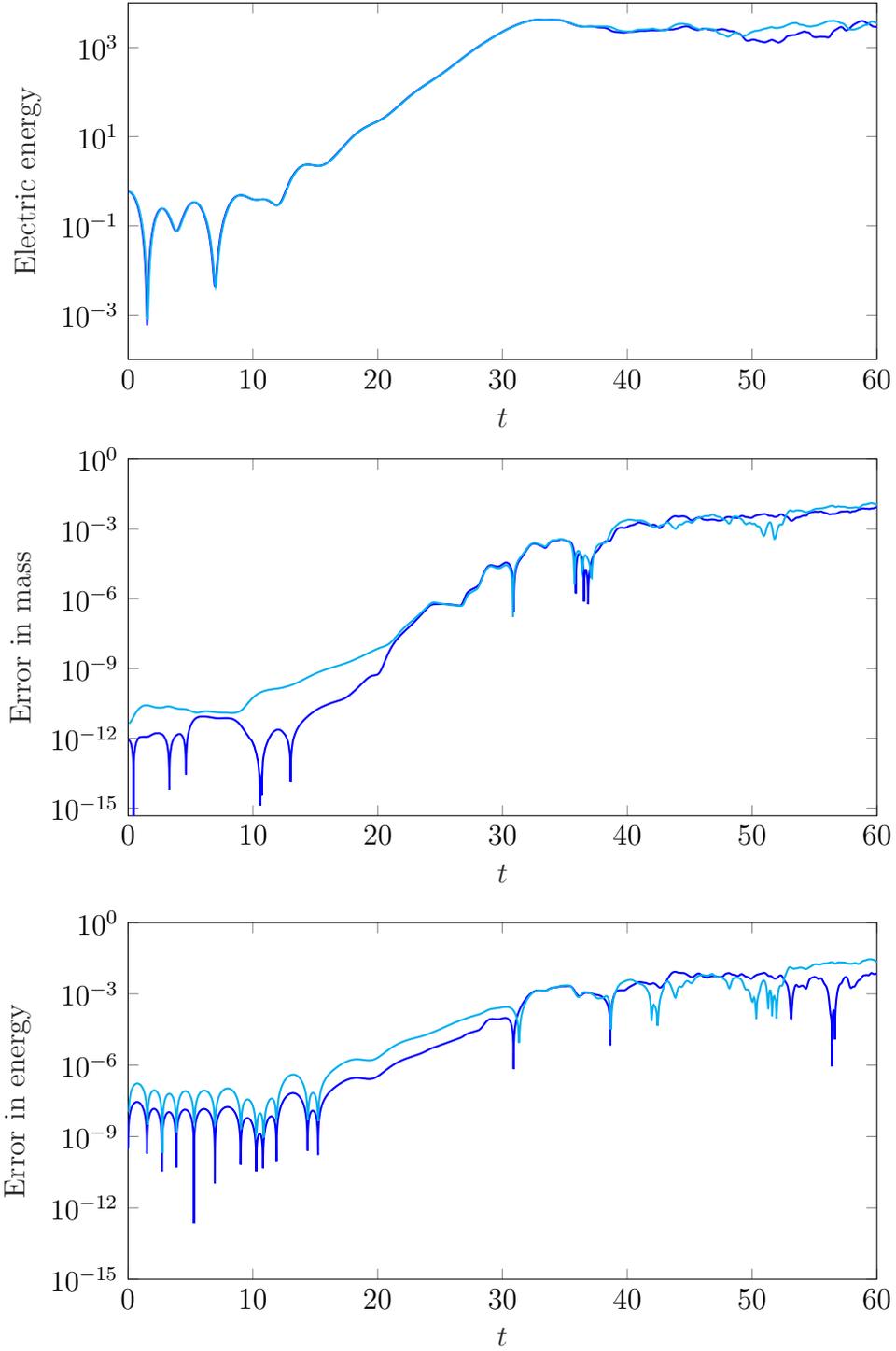}
\caption{Two stream instability simulation with $128^3$ discretization points for both
space and velocity,
final time $T=60$, rank $r=10$ and {\color{black} time step sizes $\tau=6\cdot 10^{-2}$
(cyan line) and $\tau=10^{-2}$ (blue line), see Section~\ref{sec:flagts}}.
The second order low-rank projector-splitting
algorithm~\ref{alg:secondorder} is employed. Top plot: behavior of electric energy.
Center plot: error in mass (relative). Bottom plot: error in total energy (relative).}
\label{fig:flagts}
\end{figure}

Let us perform now two simulations with the 6D two stream instability {\color{black}problem}~\eqref{eq:ts}
and $128^3$ discretization points
for both spatial and velocity variables.
The set of parameters is the same as for the example
presented in Section~\ref{sec:ordconv}, but we integrate the problem until time $T=60$ with the 
second order algorithm~\ref{alg:secondorder} {\color{black} and time step sizes 
$\tau=6\cdot 10^{-2}$ and $\tau=10^{-2}$}. The rank is fixed to
$r=10$. 
{\color{black} Again, a direct Vlasov solver would
require a total number of degrees of freedom of approximately $4.4\cdot10^{12}$, while
the dynamical low rank approach has $2\cdot 10\cdot128^3 \approx 4.2 \cdot 10^{7}$ degrees of freedom.}
As for the linear Landau problem, we investigate the behavior of the electric energy and
the conservation of mass and total energy. The results are 
summarized in Figure~\ref{fig:flagts}.
In both cases, as expected, the electric energy shows an exponential increase 
before entering into a saturation phase, and similar considerations apply also
for the error in mass and in energy {\color{black}(with slightly better results
for the simulation with lower time step size, as expected)}.
This behaviour matches well with what has been previously reported in the
literature for this problem.

\section{Performance results \label{sec:performance}}

We now investigate the performance of the low-rank projector-splitting algorithms
presented in Section~\ref{sec:matform}. This, in particular, should highlight
the efficiency of using our software framework \texttt{Ensign} and demonstrate
that GPUs provide an efficient way to run such simulations.

To perform a comparison with the GPU outcomes, we present results on a dual-socket Intel
Xeon Gold 6226R CPU based system with $2 \times 16$ CPU cores and $192$ GB of
RAM. For parallelizing the CPU code OpenMP is used and the Intel MKL library is
employed for matrix and vector operations. For the GPU performance results we
use the NVIDIA card described at the beginning of Section~\ref{sec:numerics},
and cuBLAS for BLAS operations. All simulations are conducted in double
precision arithmetics.

\subsection{CPU/GPU comparison}
We consider here the linear Landau problem~\eqref{eq:ll} discretized with $128^3$
points in both space and velocity variables.
This is done so that the effort of the $K$ and $L$ step can be directly compared.
We integrate the problem until final time $T=60$ with a {\color{black} time step}
size of $\tau=10^{-2}$ and rank $r=10$.
We report the timings (in descending order) of a single {\color{black} time step}, for the relevant
parts of the algorithms, in Table~\ref{tab:CPUGPU1} and Table~\ref{tab:CPUGPU2}
for the first order and the second order schemes, respectively.
\begin{table}
\centering
\bgroup\def\arraystretch{1.2}
\begin{tabular}{c|c|c|c}
\multicolumn{2}{c|}{CPU} & \multicolumn{2}{c}{GPU} \\
\hline
       & \textit{Wall-clock time (s)}& &\textit{Wall-clock time (s)} \\
$K$ step & $2.87\cdot 10^{0}$  & $K$ step & $2.34\cdot 10^{-2}$\\
$L$ step & $2.77\cdot 10^{0}$ & $L$ step & $2.34\cdot 10^{-2}$\\
$D$ coefficients & $1.17\cdot 10^{0}$ & $D$ coefficients & $1.13\cdot 10^{-2}$\\
$C$ coefficients & $1.17\cdot 10^{0}$ & $C$ coefficients & $1.12\cdot 10^{-2}$\\
Electric field & $9.76\cdot 10^{-2}$ & QR decomposition $K$ & $5.31\cdot 10^{-3}$\\
QR decomposition $L$ & $2.00\cdot 10^{-2}$ & QR decomposition $L$ & $5.26\cdot 10^{-3}$\\
QR decomposition $K$ & $1.96\cdot 10^{-2}$ & Electric field& $2.39\cdot 10^{-3}$\\
$S$ step & $6.72\cdot 10^{-5}$ & $S$ step & $7.43\cdot 10^{-4}$\\
\hhline{=|=|=|=}
Total & $8.12\cdot 10^{0}$ & Total & $8.30\cdot 10^{-2}$
\end{tabular}
\egroup
\caption{Breakdown of timings for a single step of the first order algorithm~\ref{alg:firstorder}, in
descending order, for CPU and GPU simulation of the linear Landau problem.
{\color{black}The number of discretization points in both space and velocity is $128^3$, the final time is 
$T=60$, the time step size is $\tau=10^{-2}$ and the rank is $r=10$.}}
\label{tab:CPUGPU1}
\end{table}
\begin{table}
\centering
\bgroup\def\arraystretch{1.2}
\begin{tabular}{c|c|c|c}
\multicolumn{2}{c|}{CPU} & \multicolumn{2}{c}{GPU} \\
\hline
       & \textit{Wall-clock time (s)}& & \textit{Wall-clock time (s)}\\
Lie splitting & $8.02\cdot 10^{0}$  & Lie splitting & $8.03\cdot 10^{-2}$\\
First $K$ step + QR & $2.83\cdot 10^{0}$ & First $K$ step + QR & $3.19\cdot 10^{-2}$\\
Second $K$ step + QR & $2.82\cdot 10^{0}$ & Second $K$ step + QR & $2.99\cdot 10^{-2}$\\
$L$ step  + QR & $2.82\cdot 10^{0}$ & $L$ step + QR & $2.89\cdot 10^{-2}$\\
$C$ coefficients & $1.12\cdot 10^{0}$ & $C$ coefficients & $1.10\cdot 10^{-2}$\\
$D$ coefficients & $1.11\cdot 10^{0}$ & $D$ coefficients & $1.09\cdot 10^{-2}$\\
Electric field & $1.01\cdot 10^{-1}$ & Electric field & $1.53\cdot 10^{-3}$\\
First $S$ step & $5.83\cdot 10^{-5}$ & First $S$ step & $6.81\cdot 10^{-4}$\\
Second $S$ step & $5.12\cdot 10^{-5}$ & Second $S$ step& $6.50\cdot 10^{-4}$ \\
\hhline{=|=|=|=}
Total & $1.88\cdot 10^{1}$ & Total & $1.96\cdot 10^{-1}$
\end{tabular}
\egroup
\caption{Breakdown of timings for a single step of the second order algorithm~\ref{alg:secondorder}, in
descending order, for CPU and GPU simulation of the linear Landau problem.
{\color{black}The number of discretization points in both space and velocity is $128^3$, the final time is 
$T=60$, the time step size is $\tau=10^{-2}$ and the rank is $r=10$.}}
\label{tab:CPUGPU2}
\end{table}

As expected, the most costly parts of the algorithms consist of the $K$ and 
the $L$ steps (with roughly the same computational time, as the degrees of freedom 
are equal in space and velocity). The cost of the $S$ step, which involves the 
solution of a problem of size $r\times r$ is negligible. The remaining 
major part of the cost lies in the computation of the $C$ and the $D$ coefficients: 
again, this is expected, as they require a matrix-matrix product
of size $N_v \times r$ (and $N_x \times r$, respectively).
A single time step of the second order scheme is, as we would expect,
approximately twice as costly as the first order scheme.

In both cases, we observe a drastic speedup (up to a factor of $100$) between
the GPU and the CPU based systems. The main reason for this is that very
efficient implementations of matrix-matrix products are available on the GPU
(in particular, in cuBLAS). This helps both in computing the coefficients as
well as performing the $K$ and $L$ step. In addition, our algorithm needs to
compute transcendental functions in order to evaluate the matrix functions in
Fourier space. This is also an area where the GPU kernels drastically outperform
the corresponding CPU implementation.

\subsection{Varying rank} \label{sec:diffrank}
We now investigate the {\color{black} performance} of the GPU implementation for
the second order low-rank projector-splitting algorithm for different ranks.
For this purpose, we consider again the 6D linear Landau problem~\eqref{eq:ll}
with $64^3$ discretization points for the spatial variables and $128^3$ discretization
points for the velocity ones.

We integrate the problem up to $T=60$ with a {\color{black} time step} size of
$\tau=10^{-2}$ and different ranks $r=5,10,15,20$. The computational times for
a single {\color{black} time step} of the simulation are reported
in {\color{black} Table}~\ref{tab:diffrankll}. First of all, we observe that 
the wall-clock time increases roughly in a linear fashion as the rank increases.
This scaling is better than the theoretical estimates
{\color{black} provided in Section~\ref{sec:soalgo}}.
\begin{table}[!htb]
\centering
\bgroup\def\arraystretch{1.2}
\begin{tabular}{c|c}
       & \textit{Wall-clock time (s)} \\
       \hline
Rank 5 & $5.12\cdot 10^{-2}$ \\
Rank 10 & $8.70\cdot 10^{-2}$ \\
Rank 15 & $1.28\cdot 10^{-1}$ \\
Rank 20 & $1.85\cdot 10^{-1}$
\end{tabular}
\egroup
\caption{Timings of a single {\color{black} time step} for the linear Landau problem with increasing ranks $r$.
{\color{black}The second order dynamical low-rank algorithm~\ref{alg:secondorder} is employed.
The number of discretization points in space is $64^3$, in velocity it is $128^3$, the final time is 
$T=60$ and the time step size is $\tau=10^{-2}$.}}
\label{tab:diffrankll}
\end{table}

\begin{figure}[!htb]
\centering
\input{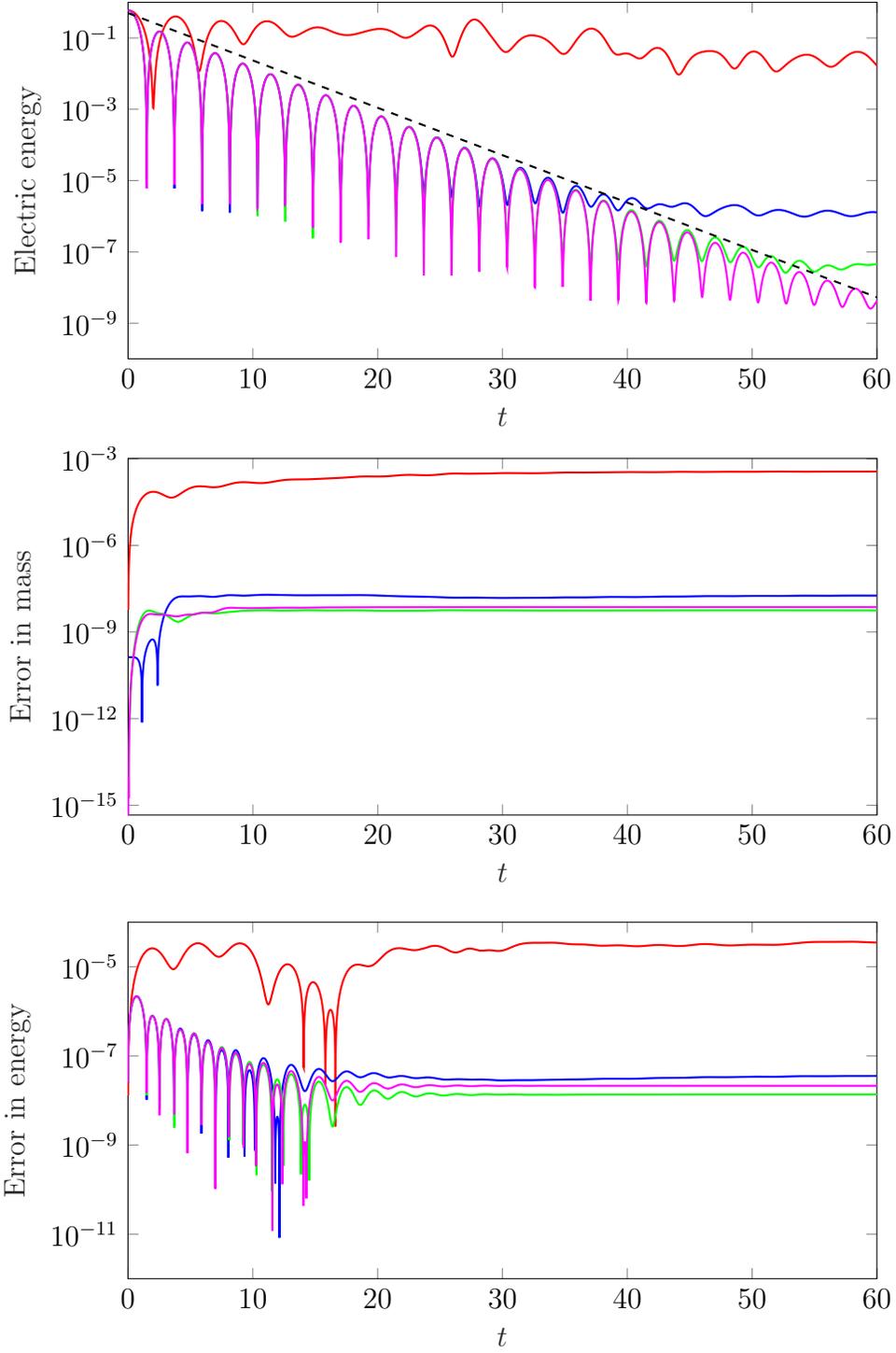}
\caption{Linear Landau simulation with $64^3$ space discretization points, $128^3$
velocity discretization points, 
{\color{black}final time $T=60$, time step} size $\tau=10^{-2}$ and
different ranks $r$, {\color{black} see Section~\ref{sec:diffrank}}.
The second order low-rank projector-splitting algorithm~\ref{alg:secondorder} is employed.
The red line corresponds to $r=5$, the blue one to $r=10$,
the green one to $r=15$ and the magenta one to $r=20$. 
Top plot: behaviour of electric energy, with reference decay rate.
Center plot: error in mass (relative). Bottom plot: error in total energy (relative).}
\label{fig:diffrankll}
\end{figure}

Then, in Figure~\ref{fig:diffrankll} we summarize the behavior of the electric energy,
of the error in mass and in total energy for all the ranks considered. We can 
clearly observe that rank 5 is not enough for the 6D problem under investigation.
Starting from rank 10, we see a substantial improvement, in particular in terms
of decay of electric energy.

\begin{figure}[!htb]
\centering
\input{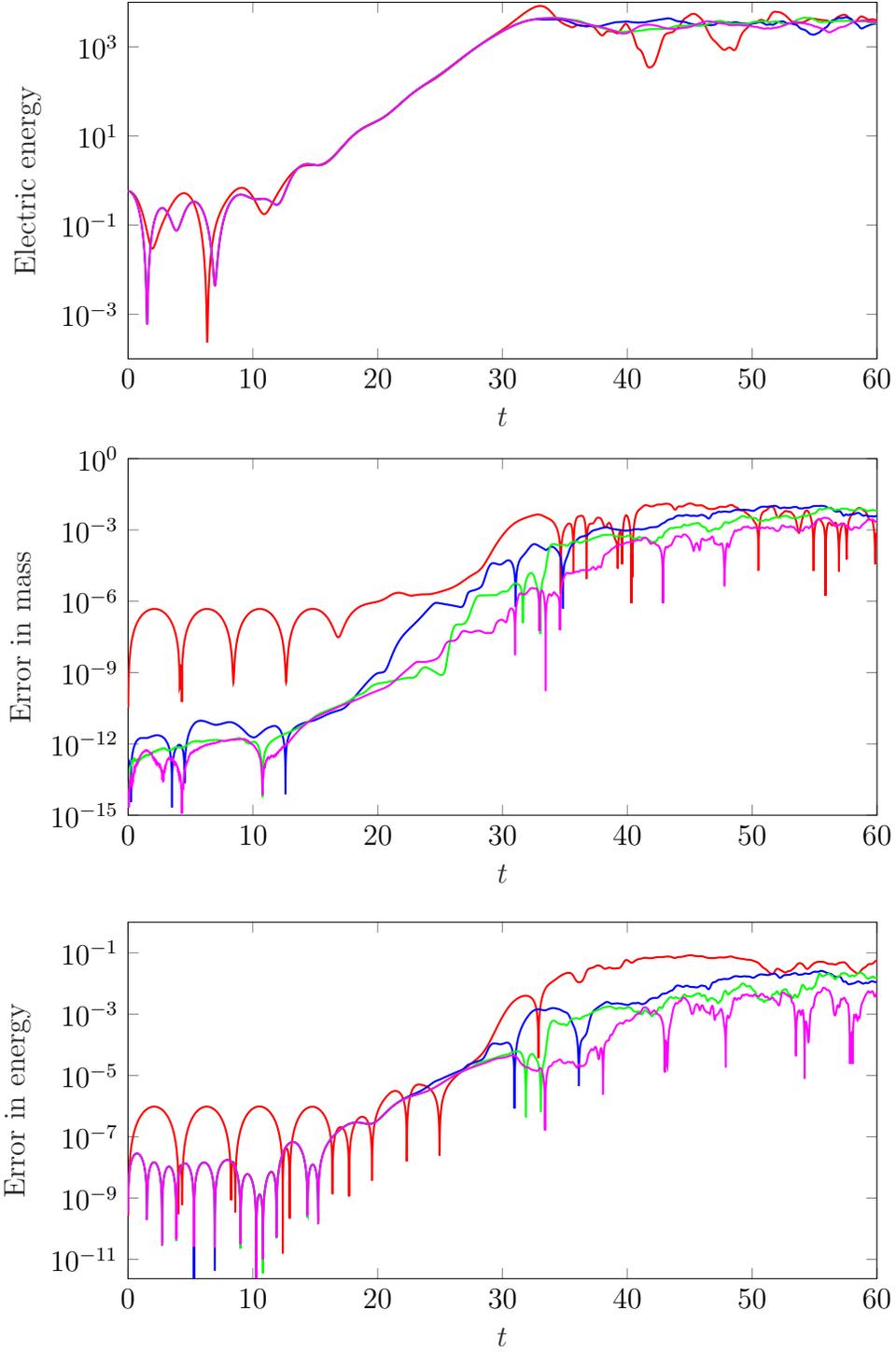}
\caption{Two stream instability simulation with $64^3$ discretization points for both
space and velocity,
{\color{black}final time $T=60$, time step} size $\tau=10^{-2}$ and
different ranks $r$, {\color{black} see Section~\ref{sec:diffrank}}.
The second order low-rank projector-splitting algorithm~\ref{alg:secondorder} is employed.
The red line corresponds to $r=5$, the blue one to $r=10$,
the green one to $r=15$ and the magenta one to $r=20$. 
Top plot: behaviour of electric energy.
Center plot: error in mass (relative). Bottom plot: error in total energy (relative).}
\label{fig:diffrankts}
\end{figure}

Finally, we repeat a similar experiment with the 6D two stream instability problem
\eqref{eq:ts}. In this case, we consider $64^3$ discretization points for 
both spatial and velocity variables. The problem is then integrated up to $T=60$ with a {\color{black} time step} size 
of $\tau=10^{-2}$ and increasing rank $r=5,10,15,20$. The computational times for 
a single {\color{black} time step} are summarized in {\color{black} Table}~\ref{tab:diffrankts}, and analogous 
conclusions as for the linear Landau simulations can be drawn.
\begin{table}
\centering
\bgroup\def\arraystretch{1.2}
\begin{tabular}{c|c}
       & \textit{Wall-clock time (s)} \\
       \hline
Rank 5 & $1.84\cdot 10^{-2}$ \\
Rank 10 & $3.06\cdot 10^{-2}$ \\
Rank 15 & $4.85\cdot 10^{-2}$ \\
Rank 20 & $7.15\cdot 10^{-2}$
\end{tabular}
\egroup
\caption{Timings of a single {\color{black} time step} for the two stream instability problem with increasing ranks $r$.
{\color{black}The second order dynamical low-rank algorithm~\ref{alg:secondorder} is employed.
The number of discretization points in both space and velocity is $64^3$, the final time is 
$T=60$ and the time step size is $\tau=10^{-2}$.}}
\label{tab:diffrankts}
\end{table}
{\color{black}In terms of electric energy, error in mass and error in total energy,
we collect the results in Figure~\ref{fig:diffrankts}}. We note that for the linear
regime rank $5$ still gives very good results in
that it perfectly predicts both the growth rate of the instability as well as
the time of its onset. Starting at saturation the rank $5$ solution tends to
overestimate the electric energy and thus the rank has to be increased. In order
to capture saturation a rank $10$ simulation is sufficient. If it is desired to
integrate far into the nonlinear regime the rank has to be increased further.

\section{Conclusions} \label{sec:conc}

We have demonstrated, by conducting numerical simulations for two test problems,
that six-dimensional Vlasov simulations using a {\color{black}projector-splitting}
dynamical low-rank algorithm
can be efficiently run on GPU based systems. In particular, we report a drastic
speedup compared to the CPU implementation, and we remark that these are the first
dynamical low-rank results obtained for the full six-dimensional problem. We
also emphasize that results with similar resolution using a direct
(Eulerian or semi-Lagrangian) Vlasov solver could only be attained on
large-scale supercomputers,
{\color{black} while our simulations have been conducted
on a single workstation equipped with an NVIDIA A100 card.}

Algorithmic efficiency has been achieved by proposing a CFL-free and second order
exponential integrator based dynamical low-rank scheme that uses a Fourier spectral
{\color{black}phase} space discretization. Implementation efficiency has been achieved by basing the
implementation on the software framework \texttt{Ensign}.

\bibliographystyle{plain}
\bibliography{dynlrbiblio}


\end{document}